\documentclass[
    ,final            
  ]
  {aipproc}

\layoutstyle{6x9}

\begin{document}

\title{Holographic Dark Energy and Present Cosmic Acceleration}

\classification{98.80.Jk}

\keywords{Cosmology, Holography, Late accelerated expansion, Dark
energy}

\author{Diego Pav\'{o}n}{
address={Departamento de F\'{\i}sica, Universidad Aut\'{o}noma de
Barcelona, 08193 Bellaterra, Spain}}

\author{Winfried Zimdahl}{
address={Institut f\"{u}r Theoretische Physik, Universt\"{a}t zu
K\"{o}ln, D-50937 K\"{o}ln, Germany}}

\begin{abstract}
We review the notion of holographic dark energy and assess its
significance in the light of the well documented cosmic
acceleration at the present time. We next propose a model of
holographic dark energy in which the infrared cutoff is set by the
Hubble scale. The model accounts for the aforesaid acceleration
and, by construction, is free of the cosmic coincidence problem.
\end{abstract}

\maketitle

\section{Introduction}
There is a growing conviction among cosmologists that the Universe
is currently experiencing a stage of accelerated expansion not
compatible with the up to now favored Einstein-de Sitter model
\cite{peebles}. According to the latter, the Universe should be
now decelerating its expansion. This conviction is deeply rooted
in observational grounds, mainly in the low brightness of high
redshift supernovae type Ia which are fainter than allowed by the
aforesaid model but consistent with accelerated models
\cite{snia}, as well as in other cosmological data. These include
the position of the first acoustic peak of cosmic microwave
background radiation (CMBR), which suggests that the Universe is
spatially flat or nearly flat \cite{cmbr}, combined with
estimations of the amount of mass at cosmological scales -see e.g.
\cite{lss}-, and correlations of the anisotropies of CMBR with
large scale structures \cite{correlations}. Overall, the data
strongly hint at a Universe dominated by some form of energy -the
so called, ``dark energy"- that would contribute about $70$
percent to the total energy density and nonrelativistic matter
(dust) which would contribute the remaining $30$ percent.

The trouble with dark energy is that we can only guess about its
nature. To begin with, it must possess a huge negative pressure,
at least high enough to violate the strong energy condition,
something required (within general relativity) to drive
accelerated expansion, and should cluster only at the highest
accessible scales. The straightforward candidate is the
cosmological constant, $\Lambda$, whose equation of state is
simply $p_{\Lambda} = -\rho_{\Lambda}$, and whose energy is evenly
distributed. Yet, it faces two serious drawbacks. On the one hand
its quantum field theoretical value is about 123 orders of
magnitude larger than observed; on the other hand, it entails the
{\em coincidence problem}, namely: ``Why are the vacuum and dust
energy densities of precisely the same order today?"
\cite{coincidence}. (Bear in mind that the energy density of dust
redshifts  with expansion as $a^{-3}$, where $a$ denotes the scale
factor of the Robertson--Walker metric). This is why a large
variety of candidates -quintessence and tachyon fields, Chaplygin
gas, phantom fields, etc.-, of varying plausibility, have been
proposed in the last years -see Ref. \cite{reviews} for reviews.
Unluckily, however, there is not a clear winner in sight.

Recently, a new form of dark energy based on the holography notion
and related to the existence of some or other cosmic horizon has
been proposed \cite{cohen}. Here, we present a specific model of
holographic dark energy that accounts for the current stage of
cosmic acceleration and is free from the coincidence problem that
besets so many models of late acceleration \cite{plb628}. The
outline of this work is as follows. We first recall the notion of
holography which is receiving growing attention and discuss
possible choices for the infrared cutoff. Then we present our
model of holographic dark energy. The last section is devoted to
the conclusions and final remarks.

\section{Holography}
We begin by recalling the notion of holography as introduced by `t
Hooft \cite{hooft} and Susskind \cite{leonard}. Consider the world
as three-dimensional lattice of spin-like degrees of freedom and
assume that the distance between every two neighboring sites is
some small length $\ell$. Each spin can be in one of two sates. In
a region of volume $L^{3}$ the number of quantum states will be
$N(L^{3}) = 2^{n}$, with $n= (L/\ell)^{3}$ the number of sites in
the volume, whence the entropy will be $S \propto (L/\ell)^{3} \ln
2$. One would expect that if the energy density does not
diverge, the maximum entropy varies as $L^{3}$, i.e., $S \sim L^{3} \,%
\Lambda^{3}$, where $\Lambda \equiv \ell^{-1}$ is to be identified
with the ultraviolet cutoff. However, the energy of most states
described by this formula would be so large that they will
collapse to a black hole of size in excess of $L^{3}$. Therefore,
a reasonable guess is that  in the quantum theory of gravity the
maximum entropy should be proportional to the area, not the
volume, of the system under consideration. (Bear in
mind that the Bekenstein--Hawking entropy is $S_{BH} = A/(4 \,%
\ell_{Pl}^{2})$, where $A$ is the area of the black hole horizon).

Consider now a system of volume $L^{3}$ of energy slightly below
that of a black hole of the same size but with entropy larger than
that of the black hole. By throwing in a very small amount of
energy a black hole would result but with smaller entropy than the
original system thus violating the second law of thermodynamics.
As a consequence, Bekenstein proposed that the maximum entropy of
the system should be proportional to its area rather than to its
volume \cite{jakob}. In keeping with this, `t Hooft conjectured
that it should be possible to describe all phenomena within a
volume by a set of degrees of freedom which reside on the surface
bounding it. The number of degrees of freedom should be not larger
than that of a two-dimensional lattice  with about one binary
degree of freedom per Planck area.

\subsection{Holographic energy interpreted as dark energy}
Inspired by these ideas, Cohen {\it et al.} \cite{cohen} argued
that an effective field theory that saturates the inequality
\\
\begin{equation}
L^{3}\, \Lambda^{3} \leq S_{BH} \, ,
\label{saturates}
\end{equation}
\\
necessarily includes many states with $R_{s} > L$, where $R_{s}$
is the Schwarzschild radius of the system under consideration.
Indeed, a conventional effective quantum field theory  is expected
to describe a system at temperature $T$ provided that $T \leq
\Lambda$. So long as $T \gg L^{-1}$, the energy and entropy will
correspond to those of radiation ($E \simeq L^{3}\, T^{4}$, and $S
\simeq L^{3}\, T^{3}$). When (\ref{saturates}) is saturated (by
setting $T = \Lambda$ in (\ref{saturates})) at $T \simeq
m_{Pl}^{2/3}\, L^{-1/3}$, the Schwarzschild radius becomes $R_{s}
\sim m_{Pl}^{2/3}\, L^{5/3} \gg L$.

Therefore it appears reasonable to propose a stronger constraint
on the infrared (IR) cutoff $L$ that excludes all states lying
within $R_{s}$, namely:
\\
\begin{equation}
L^{3}\, \Lambda^{4} \leq m_{Pl}^{2}\, L
 \label{stconstraint}
\end{equation}
\\
(obviously, $\Lambda^{4}$ is the zero--point energy density
associated to the short-distance cutoff). So, we conclude that $L%
\sim \Lambda^{-2}$ and $S_{max} \simeq S_{BH}^{3/4}$.

By saturating the inequality (\ref{stconstraint}) -which is not
compelling at all- and identifying $\Lambda^{4}$ with the
holographic dark energy density we have
\\
\begin{equation}
\rho_{x} = 3 c^{2}\, M_{p}^{2}/L^{2}\, ,
\label{rhox}
\end{equation}
\\
where $c^{2}$ is a dimensionless constant and $M_{p}^{2} \equiv
(8\pi G)^{-1}$.

\subsection{The infrared cutoff}
Before building a cosmological model of late acceleration on the
above ideas the IR cutoff must be specified. All the proposals in
the literature identify $L$ with the radius of one or another
cosmic horizon. The simplest (and most natural) choice is the
Hubble radius, $H^{-1}$. However, as shown by Hsu \cite{hsu}, this
faces the following difficulty. For an isotropic, homogeneous and
spatially flat universe dominated by nonrelativistic matter and
dark energy the Friedmann equation $\rho_{m} + \rho_{x} = 3
M_{p}^{2} \, H^{2}$ together with $\rho_{m} \propto a^{-3}$
implies that $\rho_{x}$ also redshifts as $a^{-3}$. In virtue of
the conservation equation $\dot{\rho}_{x} + 3H(\rho_{x} + p_{x}) =
0$ it follows that $p_{x}$ vanishes, i.e., there is no
acceleration. So, this first choice seems doomed.

Two other, not so natural choices, are:

\noindent $(i)$ $L = R_{ph}$ \cite{FL,raphael}, where
$$
R_{ph} = a(t)\, \int_{0}^{t}{\frac{dt'}{a(t')}}
$$
is the particle horizon. Yet, this option does not fare much
better. Assuming the dark energy to dominate the expansion,
Friedmann's equation reduces to $H\, R_{ph} = c$. Therefore, $H
\propto a^{-(1+\frac{1}{c})}$ and consequently the equation of
state parameter of the dark energy $w \equiv p_{x}/\rho_{x} =
-(1/3) +(2/3c)$ is found  to be larger than $-1/3$ whence this
dark energy candidate does not violate the strong energy condition
and cannot drive late acceleration either.

\noindent
 $(ii)$ $L = R_{H}$ \cite{Li}, where
\\
$$
R_{H} = a(t)\, \int_{t}^{\infty}{\frac{dt'}{a(t')}}
$$
\\
is the radius of the future event horizon, i.e., the boundary of
the volume a given observer may eventually see. Assuming again the
dark energy to dominate the expansion it is found that $w = -(1/3)
-(2/3c)< -1/3$. Thus, this choice is compatible with accelerated
expansion.

\section{Interacting Dark Energy}
This section focuses on our recent model of late acceleration
based on three main assumptions, namely, $(i)$ the dark energy
density is given by Eq. (\ref{rhox}), $(ii)$ $L = H^{-1}$, and
($iii$) matter and holographic dark energy do not conserve
separately but the latter decays into the former with rate $\Gamma
$, i.e.,
\\
\begin{eqnarray}
\dot{\rho}_{m} &+& 3H\rho_{m} = \Gamma\, \rho_{x}\, ,\\
\label{consv1} \dot{\rho}_{x} &+& 3H(1+w)\rho_{x} = - \Gamma
\rho_{x}\, .
 \label{consv2}
\end{eqnarray}
\\
As it can be checked, there is a relation connecting $w$ to the
ratio between the energy densities, $r \equiv \rho_{m}/\rho_{x}$,
and $\Gamma$, namely, $w = -(1+r)\Gamma/(3rH)$, such that any
decay of the dark energy $\Gamma >0$ into pressureless matter
implies a negative equation of state parameter, $w < 0$. It also
follows that the ratio of the energy densities is a constant,
$r_{0} = (1-c^{2})/c^{2}$, whatever $\Gamma$ -see Ref.
\cite{plb628} for details.

In the particular case that $\Gamma \propto H$ one has $\rho_{m}$,
$\rho_{x}$ $\propto a ^{-3m}$ and $a \propto t^{n}$ with $m =
(1+r_{0}+w)/(1+r_{0})$ and $n = 2/(3m)$. Hence, there will be
acceleration for $w < -(1+r_{0})/3$. In consequence, the
interaction is key to simultaneously solve the coincidence problem
and have late acceleration. For $\Gamma = 0$ the choice $L =%
H^{-1}$ does not lead to acceleration. Before going any further,
we wish to emphasize that models in which matter and ark energy
interact with each other are well known in the literature -see
\cite{interacting} and references therein-  and presently they are
being contrasted with cosmological data \cite{gmo}.

Obviously, prior to the current epoch of accelerated expansion
(during the radiation and matter dominated epochs) $r$ must not
have been constant but decreasing toward its current value
$r_{0}$, otherwise the standard picture of cosmic structure
formation would be irremediably spoiled (as usual, a subindex zero
means present time). To incorporate this  we must allow the
parameter $c^{2}$ to vary with time. Hence, we now have
\\
\begin{equation}
\dot{\rho}_{x}  = - 3 H \, \left[1 + \frac{w}{1 + r}\right]\,
\rho_{x}+ \frac{\left(c^{2}\right)^{\displaystyle \cdot}}{c^{2}}\,
\rho_{x}\ .
\label{dotrhox1}
\end{equation}
\\
Combining it with the conservation equation (\ref{consv2}) and
contrasting the resulting expression with the evolution equation
for $r$, namely,
\\
\begin{equation}
\dot{r} = 3 H r \left[w + \frac{1 +
r}{r}\frac{\Gamma}{3H}\right]\, ,
\label{dotr}
\end{equation}
\\
yields $\left(c^{2}\right)^{\displaystyle \cdot}/c^{2} = -%
\dot{r}/(1+r)$, whose solution is
\\
\begin{equation}
c^{2}(t) = \frac{1}{1+r(t)}\, .
\label{c2t}
\end{equation}

At sufficiently long times, $r \rightarrow r_{0}$ whence $c^{2}
\rightarrow c_{0}^{2}$.

In this scenario $w$ depends also on $c^{2}$ according to
\\
\begin{equation}
w = - \left(1 +  \frac{1}{r}\right) \left[\frac{\Gamma}{3 H} +
\frac{\left(c^{2}\right)^{\displaystyle \cdot}}{3 H c^{2}}\right]\
. \label{wx}
\end{equation}
\\
Since the holographic dark energy must fulfil the dominant energy
condition (and therefore it is not compatible with ``phantom
energy") \cite{bak}, the restriction $w \geq -1$ sets constraints
on $\Gamma$ and $c^{2}$.

\section{Discussion and conclusions}
The holographic dark energy seems to be a simple, reasonable and
elegant alternative (within general relativity) to account for the
present state of cosmic accelerated expansion. It can solve the
coincidence problem provided that matter and holographic energy do
not conserve separately. In this connection, it was pointed out by
Das {\it et al.} \cite{pierstefano}, that because the interaction
modifies the dependence of matter density on the scale factor the
observers who endeavor to fit the observational data under the
assumption of noninteracting matter will likely infer an effective
$w$ lower than $-1$. Therefore, most of the claims in favor of
phantom energy may be considered  as lending support to the dark
energy--matter interaction. Yet, models of holographic dark energy
must be further constrained by observations.

It should be noted that, contrary to what one may think, the
infrared cutoff does not necessarily change when $c^{2}$ is
varied. Indeed, the holographic bound can be expressed as
$\rho_{x}\leq 3 c^{2}\, M_{p}^{2}/L^{2}$. Now, we first considered
that it was saturated (i.e., the equality sign was assumed in the
above expression) and that $L = H^{-1}$. Since the saturation of
the bound is not at all compelling, and the ``constant" $c^2 (t)$
augments with expansion (as $r$ decreases) up to attaining the
constant value $(1+ r_{0})^{-1}$, the expression $ \rho_{x} = 3
c^2 (t) M_{p}^{2} H^{2}$, in reality, does not entail a
modification of the infrared cutoff, which  still is $L = H^{-1}$.
What happens is that, as $c^{2}(t)$ grows, the bound gets
progressively saturated up to full saturation when,
asymptotically, $c^{2}$ becomes a constant. Put another way, the
infrared cutoff stays  $L = H^{-1}$ always, what changes is the
degree of saturation of the holographic bound.

Before closing we would like to stress that there is no guarantee
that the present accelerated epoch will not be followed by
subsequent period of decelerated expansion. Models to that effect,
partly motivated by string theory demands, have been advanced,
-see, e.g. \cite{decelerated}. If, indeed, the present epoch is
followed by a decelerated one, then the future cosmic horizon will
simply not exist and models of holographic dark energy  based on
that choice of the IR cutoff will be seen as essentially flawed.

\begin{theacknowledgments}
We thank the organizers of the XXVIIIth  edition of the
``Encuentros Relativistas Espa\~{n}oles" for this opportunity. Our
research was partly supported by the Spanish Ministry of Science
and Technology under Grants BFM2003-06033 and BFM2000-1322.
\end{theacknowledgments}

\end{document}